\documentclass[aps,pra,twocolumn,showpacs,superscriptaddress]{revtex4-1}  
\usepackage{graphicx}  
\usepackage{dcolumn}   
\usepackage{bm}        
\usepackage{amssymb}   
\usepackage{color}

\newcommand{\ket}[1]{\left\vert #1 \right\rangle}
\newcommand{\bra}[1]{\left\langle #1 \right\vert}

\newcommand{\ee}{\mathrm{e}}
\newcommand{\ups}{\upsilon}
\newcommand{\lam}{\lambda}
\newcommand{\ii}{\dot{\iota}}

\newcommand{\rhoo}{\hat{\rho}}
\newcommand{\Ho}{\hat{H}}

\newcommand{\To}{\hat{T}}
\newcommand{\aoo}{\hat{a}}
\newcommand{\Ao}{\hat{A}}
\newcommand{\aod}{\hat{a}^{\dag}}
\newcommand{\Dc}{\mathcal{D}}
\newcommand{\Nc}{\mathcal{N}}

\newcommand{\pac}[1]{\left\{ #1 \right\}}
\newcommand{\pap}[1]{\left( #1 \right)}
\newcommand{\pas}[1]{\left[#1 \right]}

\newcommand{\paV}[1]{\left\Vert #1 \right\Vert}

\begin{document}

\title{Large dynamic light-matter entanglement from driving neither too fast nor too slow}

\author{O.~L. Acevedo}
\email{ol.acevedo53@uniandes.edu.co}
\author{L. Quiroga}
\author{F.~J. Rodr\'{i}guez}
\affiliation{Departamento de F\'{i}sica, Universidad de los Andes, A.A. 4976, Bogot\'{a} D. C., Colombia}
\author{N.~F. Johnson}
\affiliation{Department of Physics, University of Miami, Coral Gables, Miami, FL 33124, USA}

\begin{abstract}
A significant problem facing next-generation quantum technologies is how to generate and manipulate macroscopic entanglement in light and matter systems. Here we report a new regime of dynamical light-matter behavior in which a giant, system-wide entanglement is generated by varying the light-matter coupling at {\em intermediate} velocities. This enhancement is far larger, broader-ranged, and more experimentally accessible, than that occurring near the quantum phase transition of the same model under adiabatic conditions. By appropriate choices of the coupling within this intermediate regime, the enhanced entanglement can be made to spread system-wide or to reside in each subsystem separately.
\end{abstract}


\maketitle

\section{Introduction}

Many-body quantum dynamics lie at the core of many natural phenomena and proposed quantum technologies, including information processing through schemes such as adiabatic quantum computing \cite{Dziarmaga}. Achieving the controllable generation and manipulation of entanglement over many qubits is a key challenge, while doing so in light-matter systems is highly desirable for optoelectronic implementations. The ground state, and hence entanglement, of a quantum system can be varied in a controlled way through adiabatic perturbations, though this is in principle an infinitely slow process. Quantum Phase Transitions (QPTs) can provide a naturally occurring entangled state and it is known that the entanglement can be enhanced at the critical point \cite{BrandesENT}. Recent studies have focused on time-dependent perturbations around QPTs that are either very slow (adiabatic) \cite{PRLart} or very fast (sudden quench) \cite{Fuchs}; or small dynamic oscillations around a phase space region \cite{BastidasDicke}; or static coupling after a sudden quench \cite{Altland2,Alvermann,Alvermann1,Furuya,*Song}.

Here we consider, by contrast, the regime of intermediate perturbation velocities that has so far been overlooked. We consider the experimentally realized light-matter system of the Dicke Model (DM) \cite{Dicke}, which has been realized in a variety of systems (e.g. circuit QED \cite{Ciuti,*Marquadt} and cold atom settings \cite{Baumann,*DMEXPHamner,*DMEXPBaden,PNAS}).
We uncover a level of quantum complexity that is far richer than either the adiabatic or fast-quench regimes. The system-wide entanglement is dramatically enhanced over the static or adiabatic QPT values. Our results extend current understanding of coupled light-matter systems beyond the equilibrium ground state \cite{BrandesENT,ReslenEPL,VidalDicke,Wang2014,OlayaCastroEPL}, and also they also stand apart from more recent studies of out-of-equilibrium critical behavior  \cite{eisertnature,BastidasDicke,PRLart,Altland2,Alvermann,Alvermann1,Furuya,*Song}. Moreover, our fully quantum analysis covers all dynamical regimes from very slow adiabatic through to sudden quench,  capturing at each stage the emergent non-linear self interactions and correlations within and between each subsystem.

Our calculations employ the DM Hamiltonian \cite{Dicke}:
\begin{equation}
\hat{H} =\epsilon \hat{J}_{z}+\omega \hat{a}^{\dag }\hat{a}+ 2 \frac{\lambda (t)}{\sqrt{N}} \hat{J}_{x}\left( \hat{a} ^{\dag }+\hat{a}\right) ,
\label{eqHDick}
\end{equation}
where $N$ is the number of matter qubits, the operators $\hat{J}_{i}=\frac{1}{2}\sum_{j=1}^{N}\hat{\sigma} _{i}^{\pap{j} }$ denote collective operators of the qubits, and operator $\hat{a}^{\dag }$ ($\hat{a}$) is the creation (annihilation) operator of the radiation field. In the thermodynamic limit ($N\rightarrow \infty$), the critical value of the light-matter coupling parameter $\lambda$ is $\lambda_c = \sqrt{\omega\epsilon}/2$ while its finite-$N$ equivalent is slightly different \cite{VidalDicke,BrandesPRL,*BrandesPRE}. We treat the Hamiltonian exactly using a large basis set \cite{OscarTesis} and avoid common simplifications such as rotating-wave or semiclassical approximations. The total system evolves unitarily under Hamiltonian $\hat{H}$, starting at $t=0$ from the instantaneous ground state at $\lambda = 0$. The light-matter coupling parameter is characterized by an annealing velocity (AV) $\upsilon$ under a linear ramping $\lambda (t) =  \upsilon t$. More complicated time-dependencies can be treated but further complicate the understanding. The interval of interest in this paper is $\lambda \in [0,2]$, meaning that the driving passes across a broad range of coupling strengths below and above the QPT. While the evolution of the total system $S$ is described by a pure state $\ket{\Psi (t)}$, any subsystem $A$ is described by a density matrix $\hat{\rho}_A$ defined as the trace with respect to the other degrees of freedom not present in $A$: $ \rho_A (t) = \mathrm{tr} _{S-A} \left( \left\vert \Psi (t) \right\rangle \left\langle \Psi (t) \right\vert \right)$. Because of the global unitary condition and Schmidt decomposition, both radiation and matter have the same value of entropy $S_N$, and it provides a quantitative measure of the degree of entanglement between them \cite{Nielsen}.

This work is structured as follows: In section \ref{secmain} we present the main result of our work, namely an enhanced dynamical light-matter regime at intermediate annealing velocities. It is followed, in section \ref{secsymm}, by a deeper theoretical explanation of our findings in terms of a dynamical symmetry-breaking and effective non-linear interactions. Then, in section \ref{secpers}, we establish the prevalence of our results for a wide range of system sizes and even if the system is submitted to a dissipating environment. Finally, we provide some concluding remarks in section \ref{secconc}.

\section{Enhanced dynamical light-matter entanglement}
\label{secmain}

\begin{figure}[h]
  \includegraphics[width=0.49\textwidth]{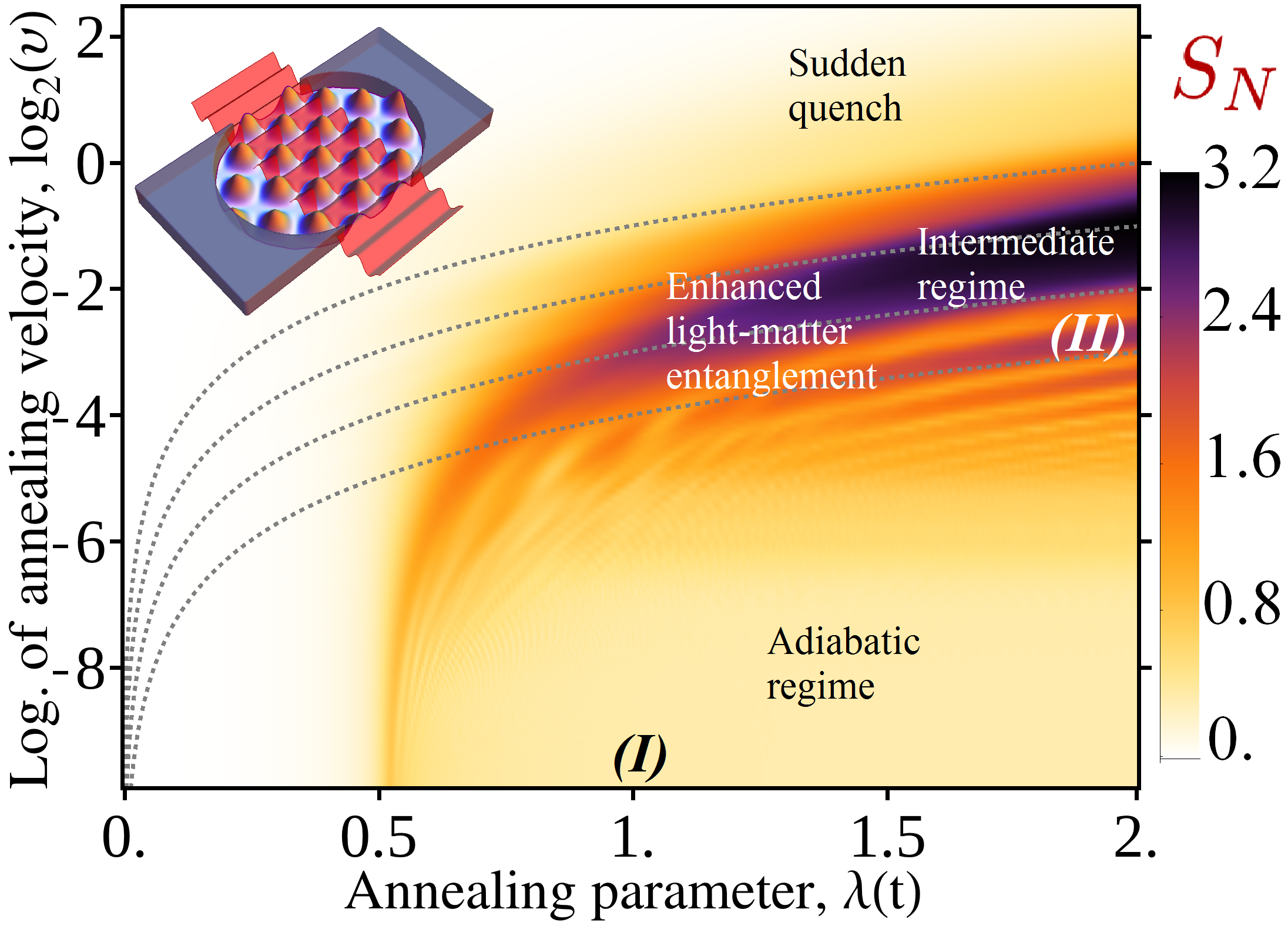}
  \caption{(color online) Dynamic evolution of the Von-Neumann entropy $S_N$   (time varies from left to right) for the Dicke Model (DM) with $N=81$ qubits. For simplicity we set $\epsilon =\omega =1$ in Eq. \ref{eqHDick} for this figure, as well as for Figs. \ref{fig3} and \ref{fig4}. The velocity range spans all regimes: from adiabatic (bottom) to sudden quench (top). Roman numerals mark where the field (Wigner) and matter (Agarwal-Wigner) distributions are depicted in Fig. \ref{fig3}. Dotted lines represent instants of equal equal time $t$ passed after initial condition (from up to down: $t=2,4,8,16$). In between those curves there is a region of novel dynamical light-matter behavior with greatly enhanced entanglement (purple) as compared to the QPT. It arises for intermediate velocities and lies well inside the coupling regime for the conventional ordered phase ($\lambda > \lambda_c = 0.5$). The QPT corresponds to $\lambda \rightarrow 0.5$ as $v\rightarrow 0$ and hence as ${\rm log} \ v\rightarrow -\infty$ (i.e. it tends toward $\lambda=0.5$ on the horizontal axis of the diagram). All dimensional quantities throughout this work are expressed in units so that energies in Eq. \ref{eqHDick} are $\omega=\epsilon=1$. Inset: a schematic representation of the DM which mimics various experimental realizations.} \label{fig1}
  \end{figure}

Figure \ref{fig1} summarizes our main finding: A novel, dynamical light-matter regime with greatly enhanced system-wide properties including entanglement, when the light-matter coupling is driven at intermediate velocities. In results of Fig. \ref{fig1} and hereinafter, we are assuming resonant energies between qubits and field mode, i.e., $\omega=\epsilon$ in Eq. \ref{eqHDick}. The peak entanglement value (purple) is far larger than the known equilibrium critical maximum \cite{BrandesENT}, i.e. much larger than what can be achieved under adiabatic conditions. Also, this enhanced entanglement extends over a far broader range, well across the $\lambda > \lambda_c$ region. As the annealing velocity increases, the critical onset point of light-matter entanglement is pushed toward larger $\lambda$ values and is no longer represented by a sharp peak, but instead a wavy plateau. At much higher velocities beyond the giant entanglement regime, $\lambda$ varies so fast that a sudden quench condition is achieved. Now the system is not quick enough to respond to the light-matter coupling, at least not in the $\lambda \in [0,2]$ interval of Fig. \ref{fig1}.

Besides exhibiting greatly enhanced values of light-matter entanglement, the intermediate regime has also some crucial practical advantages. Dotted lines in Fig. \ref{fig1} show how different are the time scales for optimal maximum light-matter entanglement in the intermediate regime as compared to the adiabatic one. In the adiabatic regime, the optimal value of light-matter entanglement is at the critical point $\lam_c$. By contrast, in the intermediate regime the optimal values are achieved well inside the $\lam>\lam_c$ phase and require evolution times that are just a tiny fraction of the time needed to reach the critical point in the adiabatic regime. In any realistic implementation, the system losses quantum information towards the environment. If these open system effects are not negligible, the intermediate regime is the only viable way to achieve optimal light-matter entanglement, before the dissipation of the environment becomes relevant.

\section{Dynamical symmetry-breaking and effective non-linear interactions}
\label{secsymm}

\begin{figure}[!h]
  \includegraphics[width=0.47\textwidth]{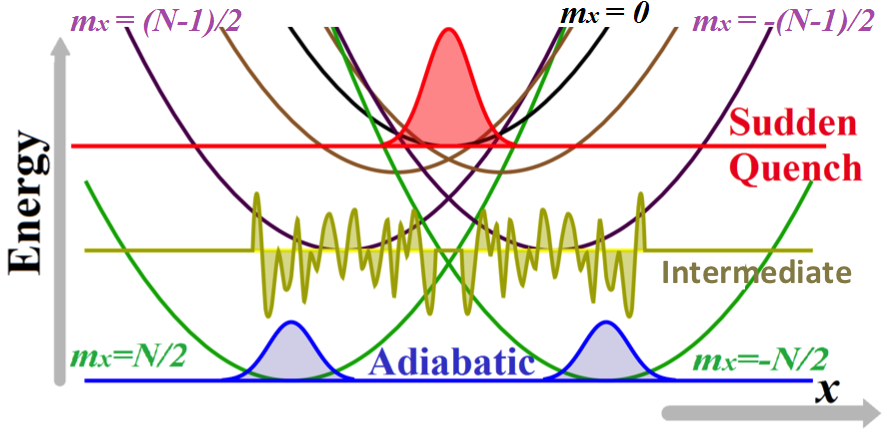}
  \caption{(color online) Schematic indicates energies of the three main outcomes when the light-matter coupling is increased at different velocities. Horizontal axis is the field position quadrature, both for sketches of the field Wigner function and for the harmonic confining potentials (parabolae). The potential felt by the field depends on the eigenvalue $m_x$ of the matter operator $\hat{J}_x$. If driving is very slow, the system stays cool in a symmetry broken ground state (blue curve) where each subsystem has entropy $\log 2$ (low matter-light entanglement).  If driving is very fast (sudden quench), the system still is in the initial $\lambda =0$ state (red curve) and light-matter entanglement is zero. For intermediate velocities, the system gets  heated such that it occupies a complicated superposition of all the  $\hat{J}_x$ eigenvalues, and hence gets disturbed by all these potentials (yellow curve).} \label{fig2}
  \end{figure}

We now develop a deeper theoretical understanding of the results in Fig. \ref{fig1}, by analyzing the underlying quantum state in the three main dynamical regimes, as illustrated in Fig. \ref{fig2}. We start by rewriting the Dicke Hamiltonian exactly as
\begin{equation}
\hat{H}=\omega \hat{b}^{\dagger}\hat{b}-\frac{4\lambda^{2}}{\omega N}\hat{J}_{x}^{2}+\epsilon \hat{J}_{z},
\end{equation}
where $\hat{b}=\hat{a}+\frac{2\lambda}{\omega\sqrt{N}}\hat{J}_{x}$. In the $\lambda > \lambda_c$ range, the last term becomes less and less relevant and the Dicke Hamiltonian can be seen as a radiation mode that feels a displaced harmonic potential whose values depend on the eigenstate $\ket{m_x}$ of $\hat{J}_{x}$ in which the matter system sits \cite{Fuchs}. Specifically, if $\lambda \gg \lambda_c$, then
\small
\begin{equation}
\hat{H}\approx\sum_{m_{x}}\left(\frac{1}{2}\omega\left[\hat{p}^{2}+\left(\hat{x}-\frac{2\lambda}{\omega\sqrt{N}}m_{x}\right)^{2}\right]
-\frac{4\lambda^{2}}{N}m_{x}^{2}\right)\ket{m_{x}}\bra{m_{x}},
\label{eqHC}
\end{equation}
\normalsize
where we have used the quadrature operators of the radiation mode. The different confining potentials depending on the eigenvalue of $\hat{J}_x$ are depicted by different parabolae in Fig. \ref{fig2}. Importantly, the energy potential is symmetrical with respect to a change in sign in $m_{x}$, which is a source of degeneracy. In addition, as $\left|m_{x}\right|$ gets bigger, the minimum value of the harmonic potential becomes lower. The ground state of this approximate Hamiltonian is any superposition of the form
\begin{equation} \ket{\psi_{0}}=\cos\theta\ket{N/2}_{m_{x}}\ket{-\beta}+\mathrm{e}^{\imath\varphi}\sin\theta\ket{-N/2}_{m_{x}}\ket{\beta},
\label{eqSSB}
\end{equation}
 i.e. it corresponds to the two minimum parabolae. Hence the symmetry of the ground state is spontaneously broken. The field state $\ket{\beta}$ is a coherent state with $\beta=\frac{2\lambda}{\omega\sqrt{N}}m_{x}$. As parity is preserved during the ramping, and adiabatic evolution keeps the energy in the lowest possible value, the projection of $\ket{\psi_{0}}$ onto the even parity sub-space ($\theta = \pi/4$ and $\varphi=0$ in Eq. \ref{eqSSB}) is the state achieved in this regime. Both the symmetry breaking and the adiabatic asymptotic value of entropy ($S_{N}=\log2$) in each subsystem can be explained by this double-well, since two coherent states are needed to describe each system.

The multiple potential picture in Fig. \ref{fig2} and Eq. \ref{eqHC} is still valid for the intermediate and sudden quench regimes. As the AV increases, the process generates a relative heating with respect to the ground state. For high enough AV (sudden quench), the system stays essentially in its starting condition, and the heating is just the consequence of the initial state being very different from the instantaneous ground state. Despite this sudden heating being very high, the simplicity of the initial state leads to no matter-light entanglement (i.e. one coherent state describes each subsystem). In the novel intermediate regime, by contrast, all confining potentials simultaneously perturb the system. A complicated superposition of non-trivial states for each parabola is generated, with complex and chaotic features. The distribution of probabilities shows a complex distribution across $J_x$ eigenvalues, leading to an entropy of each subsystem that is significantly higher in the intermediate regime than in the other two regimes. Due to its complex nature, there is no simple way to describe the structure of the dynamical state in the intermediate regime. Other forms of dynamically enhanced non-adiabatic entanglement generation are possible, including the dynamical evolution preceded by a sudden quench \cite{alba}.
\begin{figure}[!h]
  \includegraphics[width=0.49\textwidth]{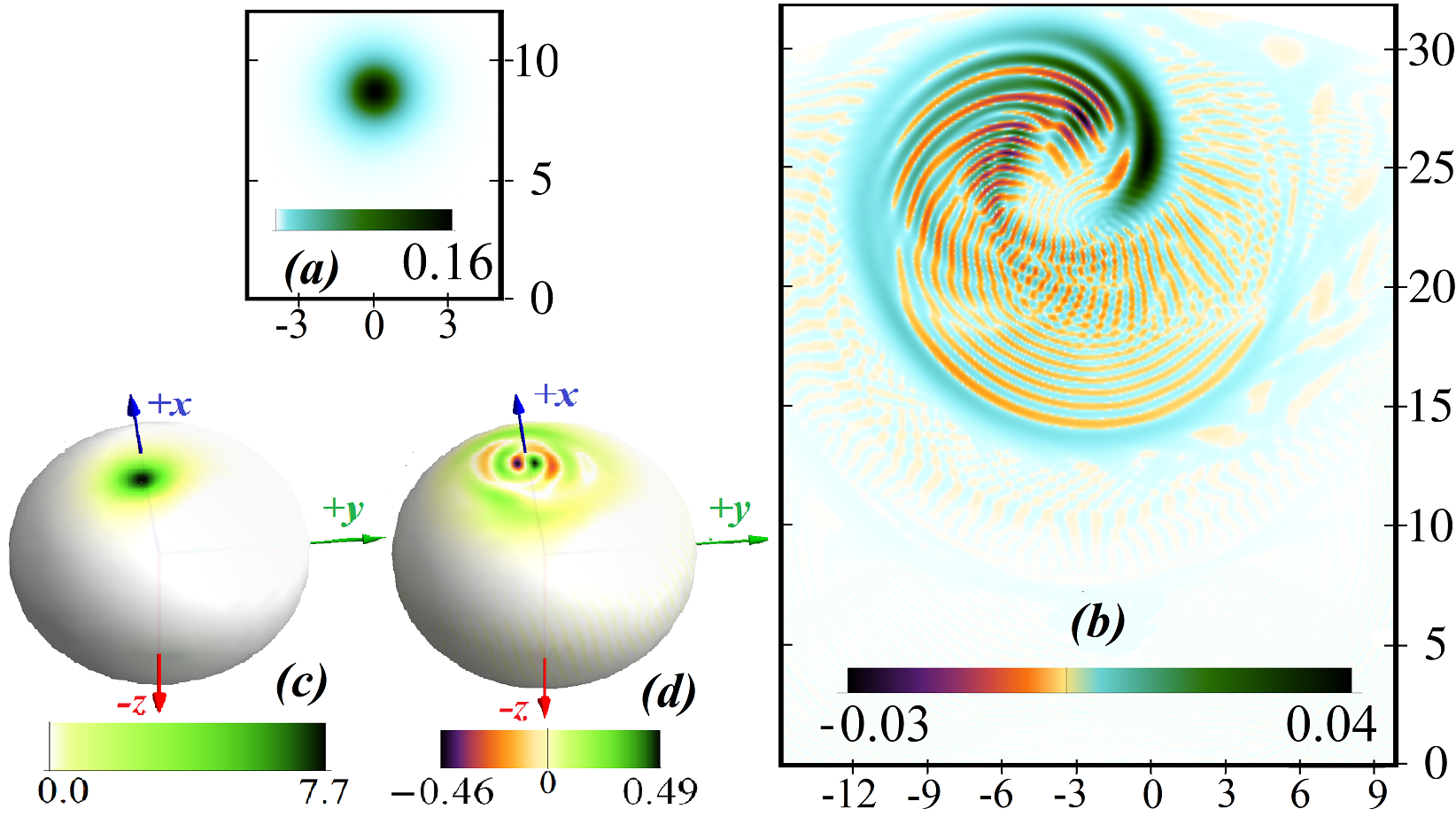}
  \caption{(color online) \textbf{(a,b)} Field Wigner distributions at the same instants as the Roman numerals marked in Fig. \ref{fig1}: (I) becomes (a) and (II) becomes (b). Horizontal scales are for $p$ momentum quadrature, while vertical ones are for the $x$ position quadrature. Both figures have same $x-p$ scale but the color scale is very different. $-x$ part of the distribution not shown because it is symmetrical due to parity: $W(x,p)=W(-x,-p)$. \textbf{(c,d)} Respective Agarwal-Wigner distributions of the qubits in the Bloch sphere. Negative regions and fragmentation (yellow curve of Fig. \ref{fig2}) are a characteristic of the giant entanglement regime. Opposite hemispheres are not shown, but can be inferred from the symmetry relation $W(\theta,\phi+\pi)=W(\theta,\phi)$ due to conserved parity.} \label{fig3}
  \end{figure}

 The distinct behavior in the intermediate regime can also be interpreted in terms of the realization of non-linear self-interactions in each subsystem \cite{ReslenEPL,OlayaCastroEPL}, with qubits subjected to One-Axis Spin Squeezing with a Transverse Field \cite{QuaSpiSqu,Transverse}. As the light-matter coupling increases in time from zero into the $\lambda > \lambda_c$ range, the system moves out of its frozen initial state. The interaction term in the Dicke Hamiltonian begins to dominate and each subsystem works as a mediator of the other's self-interaction. Matter-light interaction is in principle linear and generates little entanglement in the adiabatic limit, since the system has enough time to continuously stabilize in response to the perturbation. The intermediate regime is the only one in which non-linearities can develop significantly. High values of squeezing in both subsystems are then generated \cite{OscarTesis}. However, this process does not last indefinitely: each subsystem starts to retain quantum information of the other, which is the moment when the entropy grows. The effective interaction is broken, leaving the qubits and radiation mode entangled with each other but not within themselves. Despite being a single radiation mode, the field acts as a reservoir that dissipates the quantum correlations present in the squeezed states of each subsystem. The quantum correlations developing in the field system can be represented by the Wigner quasi-probability distribution in Figs. \ref{fig3}(a) and (b). See appendix \ref{appwig} for definitions of these distributions. This shows that the distribution becomes highly fragmented yet retains some order, reflecting the complexity of the light subsystem in the intermediate regime. Round-tailed interference patterns as in Fig. \ref{fig3} have been obtained in light with a non-linear Kerr-like interaction following a Fokker-Planck equation \cite{Milburn1}. This confirms that the field experiences an effective non-linear interaction. Similar signatures of complexity arise in the matter subsystem, specifically the matter density matrices, as shown in the spherical Agarwal-Wigner functions in Figs. \ref{fig3}(c) and (d) \cite{WignerQubit1}.

\begin{figure}[!h]
  \includegraphics[width=0.44\textwidth]{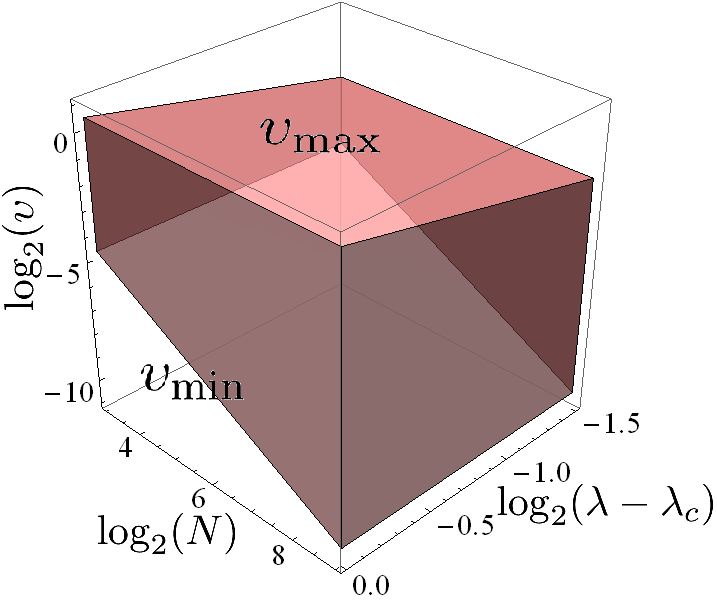}
  \caption{(color online) Dynamical phase diagram showing the new enhanced entanglement regime (shaded red) defined by instances where $S_N > \log2$ since $S_{N}\rightarrow \log 2$ is the asymptotic value for the equilibrium ordered phase. For visual clarity in this figure, we take $S_N > \log2+0.05$. Below a minimum annealing velocity $\upsilon_{\min}$ the system's behavior is adiabatic, while for $\upsilon>\upsilon_{\max}$ it corresponds to sudden quench. The adiabatic boundary depends on system size as $\upsilon_{\min} \propto N^{-1}$. The sudden quench boundary is size-independent, but depends on the value of $\lambda_d$ that is reached during the driving:  $\upsilon_{\max} \propto (\lambda_d-\lambda_c)^{3/2}$. See appendix \ref{appphas} for more details.} \label{fig4}
  \end{figure}

\section{Persistence of results against dissipation and different system sizes}
\label{secpers}

Figure \ref{fig4} demonstrates how the range of velocities that classify as `intermediate' actually increases with increasing number of qubits $N$, meaning that the enhanced entanglement regime (EER) begins to dominate the space of behaviors as opposed to becoming a small niche. The EER can be imagined as lying between a lower bound AV $\upsilon_{\min}$ which marks the adiabatic evolution, and an upper bound one $\upsilon_{\max}$ defining the AV at which the sudden quench approximation starts to be valid. Specifically, Fig. \ref{fig4} shows the dynamical phase diagram of the intermediate regime in which the giant entanglement occurs, including its scaling behavior (dependence on system size $N$). The adiabatic evolution is more difficult to achieve as the number of atoms increases. The other main variable is the value of $\lambda_d$ reached by the annealing. The sudden quench condition requires higher AVs as this $\lambda_d$ gets bigger. The oscillatory behavior near the adiabatic regime has been smoothed out in order to make the phase boundary visually clearer. More details on how this diagram was obtained are given in appendix \ref{appphas}.

In section \ref{secmain}, we used Von Neumann entropy $S_N$ as a measure of light-matter entanglement in the case that the entire system is pure. However, in order to have an idea of the decoherence effects of a leaky cavity we have to analyze the DM as an open system, and then, $S_N$ is no longer a good entanglement witness. Instead, we use quantum negativity, whose non-zero value is a sufficient condition for bipartite entanglement in the open light-matter system \cite{Vidalneg}. Quantum negativity is defined as,
 \begin{equation}
 \Nc \pap{\rhoo} = \frac{\paV{\rhoo^{\Gamma_q}}_{1}-1}{2} ,
 \label{eqneg}
 \end{equation}
where $\rhoo^{\Gamma_q}$ is the partial transpose of $\rhoo$ with respect to the matter subsystem, and $\paV{\Ao}_{1}\equiv\mathrm{tr}\pac{\sqrt{\Ao^{\dagger}\Ao}}$ is the trace norm.
\begin{figure}[!h]
  \includegraphics[width=0.49\textwidth]{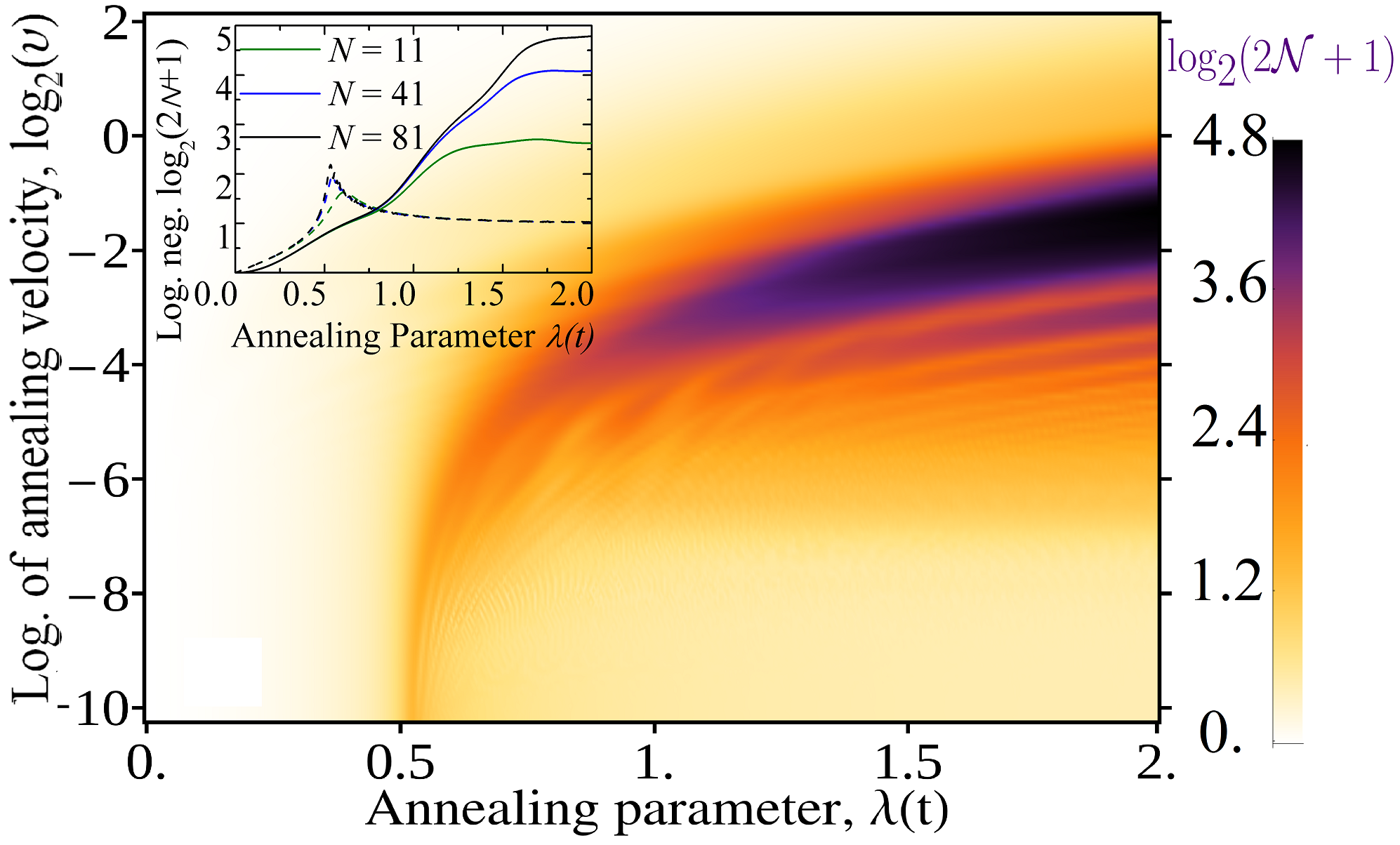}
  \caption{(color online) Dynamical profile analogous to Fig. \ref{fig1} but now for logarithmic negativity $\log_2\pap{2 \Nc +1 }$. There is strong resemblance between both profiles. This justifies the use of $S_N$ as the entanglement witness for the closed system case while changing to $\log_2\pap{2 \Nc +1 }$ when the system is open. The inset shows the dynamical evolution of logarithmic negativity for a near-adiabatic ramping velocity $\log_2(\ups)=-8.95$ (dashed lines) and an intermediate one $\log_2(\ups)=-1.55$ (solid lines). Different line colors represent different system sizes. This inset shows that even with relative small system sizes, $\log_2\pap{2 \Nc +1 }$ has qualitatively similar behaviors for the same $\ups$, so that conclusions about the robustness of the dynamic light-matter entanglement against decoherence can be extrapolated to bigger $N$.}
  \label{fig5}
  \end{figure}
However, one may wonder if switching from one form of entanglement measure to the other has any justification. Figure \ref{fig5} shows a dynamical profile of the ramping process analogous to that of Fig. \ref{fig1} (same parameters, unitary evolution, and system size $N=81$), but with logarithmic negativity $\log_2\pap{2\Nc+1}$. The resemblance with the $S_N$ graphic is quite apparent, and the EER is again clearly noticeable. This should be no surprise as both measures virtually codified the same information. If total $\rhoo$ is a pure state, and $\pac{p_i}$ is the spectrum of $\rhoo_q$ in that case (which is the same as the spectrum of $\rho_b$ because of Schmidt decomposition), both negativity $\Nc$ and $S_N$ can be written in terms of that spectrum. The first one would be $2 \Nc +1= \sum_{i,j} \sqrt{p_i p_j}$, while the second one is $S_N = -\sum_i p_i \log p_i$. Therefore, Fig. \ref{fig5} confirms that both entanglement measures are well connected and suitable for witnessing light-matter entanglement. The reason why $S_N$ was preferred in section \ref{secmain} is because its wider usage and well established connection to other quantum information concepts \cite{Nielsen}.

\begin{figure}[!h]
  \includegraphics[width=0.49\textwidth]{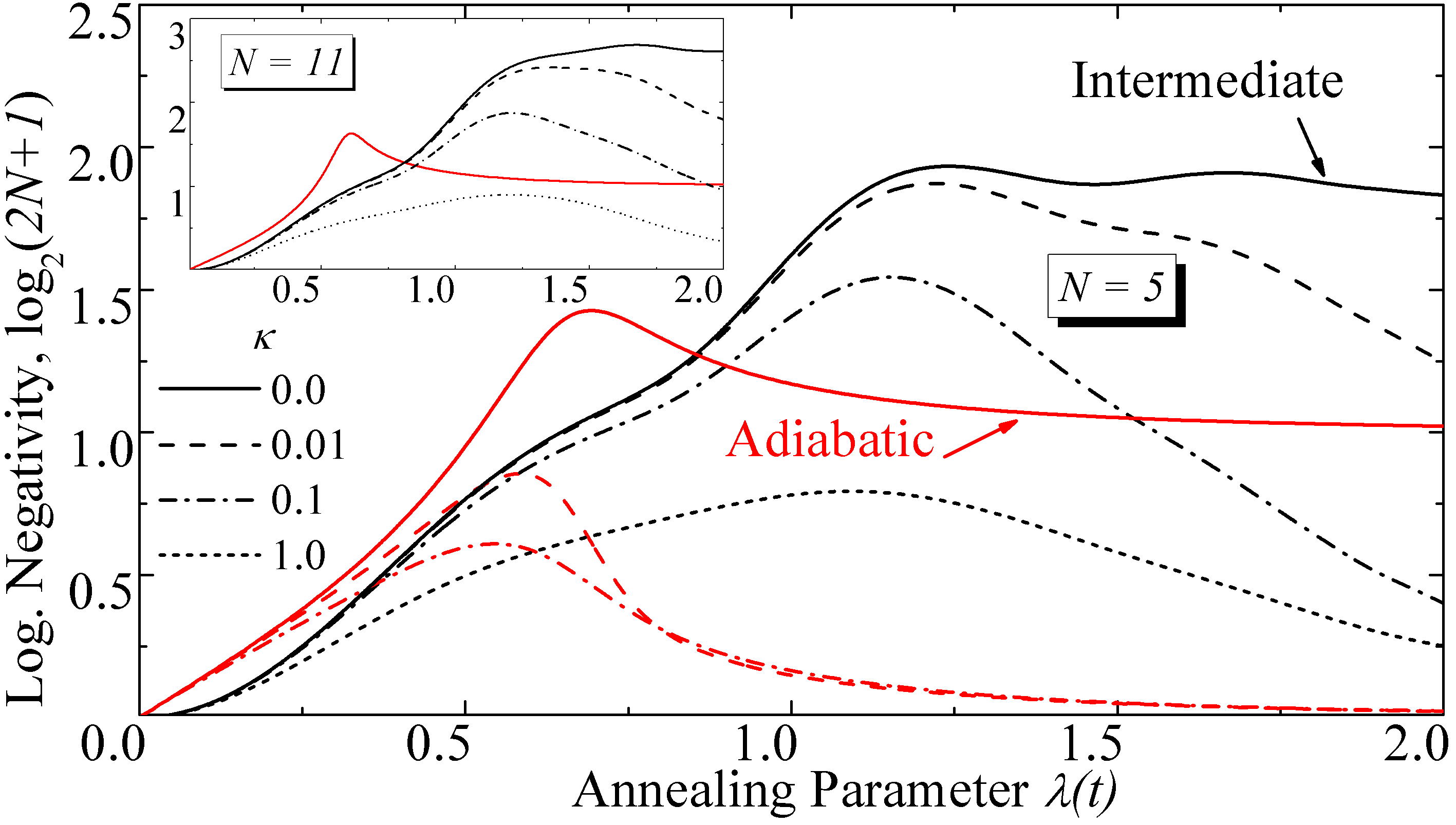}
  \caption{(color online) Effect of cavity losses on light-matter entanglement generation, witnessed by quantum logarithmic negativity. Cavity's field decay rate is $\kappa$ (different line style). Black curves depict giant entanglement regime $\log_2(\ups)=-1.58$, while red curves represent adiabatic one $\log_2(\ups)=-8.96$. Main figure is for $N=5$ system's size while inset represents analogous $N=11$ results. Dynamical entanglement regime is bigger and more robust against open system losses as $N$ increases. In opposition, near-adiabatic entanglement vanishes even for weak losses since annealing time is too long as compared to decoherence time. Zero temperature has been assumed and $\kappa$ is measured in units of field free frequency $\omega$. Finite temperature cases have very similar tendencies.} \label{fig6}
  \end{figure}

It should be stressed that the giant entanglement regime, central to present work, is accessible under current experiments very similar to the sketch in inset of Fig. \ref{fig1} \cite{Baumann,*DMEXPHamner,*DMEXPBaden,PNAS}. In particular, Klinder et al. have been able to ramp a cold atom Bose-Einstein condensate whose dynamics are governed by the DM, and with annealing velocities corresponding very well to the giant entanglement regime predicted by our analysis \cite{PNAS}. Moreover, cavity decay rates, which measure the importance of losses, are an order of magnitude lower than the main evolution energy $\omega$, and can be well simulated by open cavity master equations \cite{PNAS}. Our analysis including losses predicts the survival of giant light-matter entanglement in those experiments -- indeed it occurs even when open system losses are present, as shown in Fig. \ref{fig6}. We use quantum negativity as the light-matter entanglement measure for the open system case \cite{Vidalneg} since $S_N$ will no longer be a good bipartite entanglement measure. Details of the definition of quantum negativity and the open system solution, are given in the appendix \ref{appopen}. Since the light-matter entanglement increases as $N$ increases, we predict that  experimental system sizes $N\approx 10^5$ will develop significantly more robust light-matter entanglement against losses than for small systems. By contrast, entanglement generation in the well-studied near-adiabatic regime cannot be achieved under current experimental setups, since decoherence times are much shorter than the annealing times required by this regime.

We have then established the persistence of the EER for a wide range of system sizes and even under the effect of dissipation effects. A remnant concern could be that the system sizes $N$ accessible to open system numerical solution in the present work are well smaller than the ones examined in the pure case. This could cast doubt on whether the conclusions brought from this small size result have any general validity. The inset of figure \ref{fig5} shows that despite even at $N=11$, there are good signatures of the dynamical phase with giant light-matter entanglement, and that it only gets stronger as $N$ increases. This means that, provided the open system's parameters like $\kappa$ and $\bar{n}$ remain the same,  any deleterious effect caused by the leakage of quantum information to the environment is only weaker with sizes of the order $N=81$ than the ones numerical accessible.

\section{Conclusions}
\label{secconc}

We have unraveled a previously unnoticed dynamical regime where enhanced light-matter entanglement arises at intermediate ramping velocities. We have provided theoretical insights of this phenomenon by means of a dynamical symmetry breaking and effective non-linear self-interactions. We have further argued that our main results are still present under more realistic settings, as those already accessible by current experiments, with bigger system sizes and a dissipating environment. Finally, we note that by moving around the parameter space in time in Fig. \ref{fig1}, the enhanced squeezing and entanglement can be altered within the matter and light subsystems separately, and then transferred by means of the light-matter coupling. Potential applications include high precision quantum metrology and a range of quantum information processing technologies \cite{Metrology,*MetrologyOPT,*Rey2007,QuInCo,VedralE}.

\begin{acknowledgements}
O.L.A, L.Q. and F.J.R. acknowledge financial support from Proyectos Semilla-Facultad de Ciencias at
Universidad de los Andes (2010-2014) and project {\it Quantum control of non-equilibrium hybrid
systems}, UniAndes-2013. O.~L.~A. acknowledges financial support from Colciencias, Convocatoria 511.
\end{acknowledgements}

\appendix

\section{Subsystem quasi-probabilities distributions}
\label{appwig}

In order to provide a visual depiction of both matter and field subsystems in the main text, we employed phase space representations of the density matrix by means of Wigner functions. The matter states were represented by the Agarwal-Wigner function (AWF), which is a Bloch sphere representation of the subsystem's density matrix $\rhoo_q$ \cite{WignerQubit1},
 \begin{equation}
 W_q(\theta,\phi)=\sum_{l=0}^{N}\sum_{m=-l}^{l} T_{l,m}Y_{l,m}(\theta,\phi),
 \label{eqWignerqub}
 \end{equation}
where $Y_{l,m}$ are the spherical harmonics; and terms $T_{l,m}=\mathrm{tr}\pac{\rhoo_q \To_{l,m}}$ are expected values of the multipole operator,
\begin{eqnarray}
\hat{T}_{l,m}=\sum_{M,M^{\prime}=-j}^{j} (-1)^{j-m} \sqrt{2l+1} \nonumber \\
 \pap{\begin{array}{ccc} j & l & j \\ -M  & m & M^{\prime} \end{array}} \ket{jM}\bra{jM^{\prime}},
\end{eqnarray}
where $j=N/2$, and ${\tiny \pap{\begin{array}{ccc}  j & l & j \\ -M  & m & M^{\prime} \end{array}} }$ is the Wigner $3j$ symbol.\\

The respective Wigner function for the field density matrix $\rhoo_b$ is \cite{quantumoptics,Fuchs},
\begin{equation}
W_b\pap{ \alpha ,\rhoo _b } = \sum_{n=0}^{\infty }\left( -1\right) ^{n}\left\langle
n\right\vert \hat{D}^{\dag }\left( \alpha \right) \rhoo_b \hat{D}\left( \alpha \right)
\left\vert n\right\rangle, \label{eqWignerbos}
\end{equation}
where $\hat{D}\left( \alpha \right) =\mathrm{e}^{\alpha \aoo^{\dag }-\alpha ^{\ast }\aoo}$ is the displacement operator, and $\alpha \in \mathbb{C}$. The displacement parameter $\alpha$ can be expressed in terms of the field's position ($x$) and momentum ($p$) quadratures as its real and imaginary part, $\sqrt{2}\alpha=x+\ii p$.

\begin{figure}[!h]
  \includegraphics[width=0.49\textwidth]{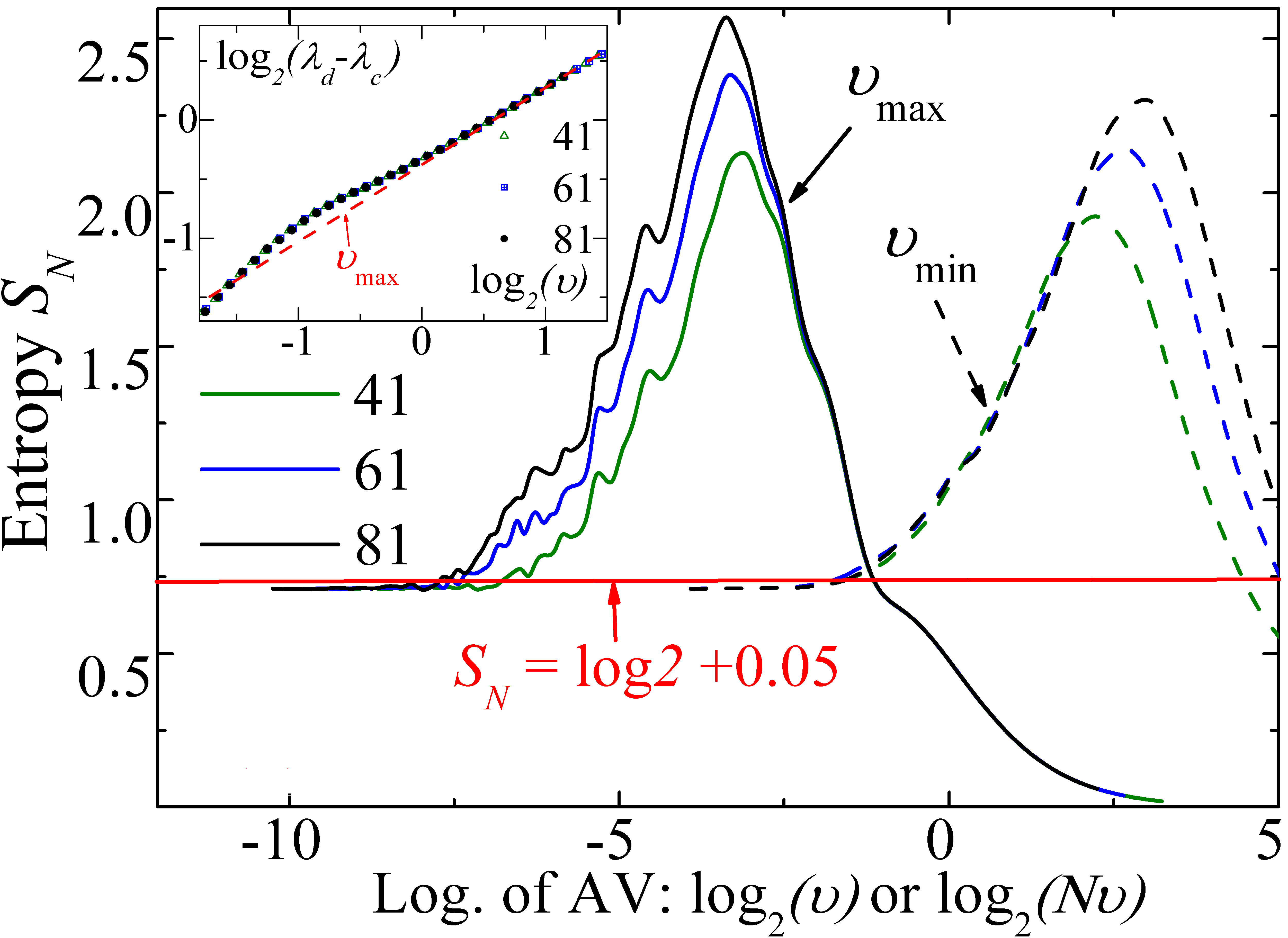}
  \caption{(color online) Numerical evidence for constructing the dynamical phase diagram of Fig. \ref{fig4} of main text. Solid (dashed) curves depict the value of $S_N$ as a function of $\log_2\pap{\ups}$ ($\log_2\pap{N\ups}$) when $\lam = 2.0$ for different system sizes. Point $\ups_{\mathrm{max}}$  ($N\ups_{\mathrm{min}}$) is obtained in a size-independent way where curves touch the red horizontal line, which is the defining value for the dynamical phase boundary. The inset shows the dependence of point $\ups_{\mathrm{max}}$ on the value of $\lam_d$ reached, and its fitting to the $\upsilon_{\max} \propto (\lambda-\lambda_c)^{3/2}$ dependence (red dashed line). In opposition, the point $N\ups_{\mathrm{min}}$ is independent of the value of $\lam_d$ reached. In principle, solid and dashed curves represent the same data, but dashed curves have been smoothed in order to establish a cleaner boundary for $\ups_{\mathrm{min}}$.}
  \label{figureapp}
  \end{figure}

\section{Numerical evidence of phase diagram}
\label{appphas}

This section provides a detailed explanation on how we obtained the phase diagram of the EER depicted in Fig. \ref{fig4} of main text. The EER was defined as the dynamical region in Fig. \ref{fig1} of main text where entropy is bigger than $\log 2+0.05$ well inside the ordered $\lam > \lam_c$ phase. This region can be seen as bounded by a maximum AV $\ups_{\mathrm{max}}$ over which the sudden quench approximation is valid, and a minimum AV $\ups_{\mathrm{min}}$ below which the adiabatic condition is fulfilled.

From Fig. \ref{fig1} of main text, it is clear that $\ups_{\mathrm{min}}$ does not depend on the value $\lam_d$ of the annealing parameter reached (the lower boundary forms an horizontal line in the figure). This is expected as the ground state in all the ordered phase has an asymptotic value of $S_N \rightarrow \log 2$, and the adiabatic condition should only depend on the system size $N$. The scaling $\ups_{\mathrm{min}} \propto N^{-1}$ comes from a well established relation of the minimal energy gap $\Delta$ at the critical threshold \cite{BrandesPRE,PRLart}, and it is confirmed in the dashed curves of Fig. \ref{figureapp} where using this kind of scaling of the AV makes the lower boundary of the EER size independent.The upper bound, $\ups_{\mathrm{max}}$ depends on $\lam_d$ and the inset of Fig. \ref{figureapp} shows that the fitting $\ups_{\mathrm{max}} \propto (\lambda_d-\lambda_c)^{3/2}$ is a good approximation. On the other, it is clear that $\ups_{\max}$ does not depend on system size, as all three solid curves of Fig. \ref{figureapp} reach the red line at the same point. The relations $\ups_{\mathrm{max}} \propto (\lambda_d-\lambda_c)^{3/2}$ and $\ups_{\mathrm{min}} \propto N^{-1}$ with the addition of one particular point (as the ones obtained in Fig. \ref{figureapp}) is all that is needed to produce Fig. 3 in main text.

\section{Open System Evolution}
\label{appopen}

The composed DM light-matter density matrix $\rhoo (t)$ evolves under a unitary part generated by $\Ho$ (see Eq. 1 of main text) and a dissipative part due to a field lossy cavity. The evolution of $\rhoo (t)$ is modeled by a Master Equation in Lindblad form \cite{breuer},
\begin{equation}
\frac{d}{d t}\rhoo = -\ii \pas{\Ho,\rhoo} +2\kappa \pap{\bar{n}+1} \Dc \pap{\rhoo;\aoo}+2\kappa \bar{n} \Dc \pap{\rhoo;\aod},
\label{eqME}
\end{equation}
 where $\kappa$ is the damping rate of the cavity, and $\bar{n}$ is the thermal mean photon number. Also, for any operator $\Ao$, the Lindblad superoperator $\Dc$ is defined as ($\pac{,}$ denotes anti-commutator),
\begin{equation}
\Dc \pap{\rhoo;\Ao}=\Ao \rhoo \Ao^{\dagger}-\frac{1}{2}\pac{\Ao^{\dagger}\Ao,\rhoo}.
\label{eqDcal}
\end{equation}
The general initial condition (when $\lambda(0) =0$) consist of an unentangled light-matter state,
\begin{equation}
\rhoo (0) = \ket{-N/2}_z\bra{-N/2}_z \otimes \frac{\ee^{-\beta \aod\aoo}}{\mathrm{tr}\pac{\ee^{-\beta \aod\aoo}}},
\label{eqinitrhoo}
\end{equation}
where $\ee^{-\beta} = \bar{n}/\pap{\bar{n}+1}$, and $\beta=1/T$ is the inverse temperature in natural units. This condition means zero excitations in the matter state and field thermalization. The field state is in fact the steady equilibrium state for the respective terms in Eq. \ref{eqME}.

We have obtained exact numerical solutions of Eq. \ref{eqME}, either in the pure DM Hilbert state when $\kappa =0$, or in the space of density matrices otherwise, using the DM Hamiltonian or the Liouvillian superoperator as the respective dynamical generator \cite{breuer}. Since the dimension of the numerical evolution vector space is squared as soon as the unitary condition is broken, only relatively small values of $N$ can be investigated under open conditions. As the field Hilbert space is infinite dimensional, we have truncated its Fock basis $\pac{\ket{n}}$ up to a number where numerical results converge.

\bibliography{References}

\begin{thebibliography}{39}%
\makeatletter
\providecommand \@ifxundefined [1]{%
 \@ifx{#1\undefined}
}%
\providecommand \@ifnum [1]{%
 \ifnum #1\expandafter \@firstoftwo
 \else \expandafter \@secondoftwo
 \fi
}%
\providecommand \@ifx [1]{%
 \ifx #1\expandafter \@firstoftwo
 \else \expandafter \@secondoftwo
 \fi
}%
\providecommand \natexlab [1]{#1}%
\providecommand \enquote  [1]{``#1''}%
\providecommand \bibnamefont  [1]{#1}%
\providecommand \bibfnamefont [1]{#1}%
\providecommand \citenamefont [1]{#1}%
\providecommand \href@noop [0]{\@secondoftwo}%
\providecommand \href [0]{\begingroup \@sanitize@url \@href}%
\providecommand \@href[1]{\@@startlink{#1}\@@href}%
\providecommand \@@href[1]{\endgroup#1\@@endlink}%
\providecommand \@sanitize@url [0]{\catcode `\\12\catcode `\$12\catcode
  `\&12\catcode `\#12\catcode `\^12\catcode `\_12\catcode `\%12\relax}%
\providecommand \@@startlink[1]{}%
\providecommand \@@endlink[0]{}%
\providecommand \url  [0]{\begingroup\@sanitize@url \@url }%
\providecommand \@url [1]{\endgroup\@href {#1}{\urlprefix }}%
\providecommand \urlprefix  [0]{URL }%
\providecommand \Eprint [0]{\href }%
\providecommand \doibase [0]{http://dx.doi.org/}%
\providecommand \selectlanguage [0]{\@gobble}%
\providecommand \bibinfo  [0]{\@secondoftwo}%
\providecommand \bibfield  [0]{\@secondoftwo}%
\providecommand \translation [1]{[#1]}%
\providecommand \BibitemOpen [0]{}%
\providecommand \bibitemStop [0]{}%
\providecommand \bibitemNoStop [0]{.\EOS\space}%
\providecommand \EOS [0]{\spacefactor3000\relax}%
\providecommand \BibitemShut  [1]{\csname bibitem#1\endcsname}%
\let\auto@bib@innerbib\@empty
\bibitem [{\citenamefont {Dziarmaga}(2010)}]{Dziarmaga}%
  \BibitemOpen
  \bibfield  {author} {\bibinfo {author} {\bibfnamefont {J.}~\bibnamefont
  {Dziarmaga}},\ }\href {\doibase 10.1080/00018732.2010.514702} {\bibfield
  {journal} {\bibinfo  {journal} {Adv. Phys.}\ }\textbf {\bibinfo {volume}
  {59}},\ \bibinfo {pages} {1063} (\bibinfo {year} {2010})}\BibitemShut
  {NoStop}%
\bibitem [{\citenamefont {Lambert}\ \emph {et~al.}(2004)\citenamefont
  {Lambert}, \citenamefont {Emary},\ and\ \citenamefont
  {Brandes}}]{BrandesENT}%
  \BibitemOpen
  \bibfield  {author} {\bibinfo {author} {\bibfnamefont {N.}~\bibnamefont
  {Lambert}}, \bibinfo {author} {\bibfnamefont {C.}~\bibnamefont {Emary}}, \
  and\ \bibinfo {author} {\bibfnamefont {T.}~\bibnamefont {Brandes}},\ }\href
  {\doibase 10.1103/physrevlett.92.073602} {\bibfield  {journal} {\bibinfo
  {journal} {Phys. Rev. Lett.}\ }\textbf {\bibinfo {volume} {92}},\ \bibinfo
  {pages} {073602} (\bibinfo {year} {2004})}\BibitemShut {NoStop}%
\bibitem [{\citenamefont {Acevedo}\ \emph {et~al.}(2014)\citenamefont
  {Acevedo}, \citenamefont {Quiroga}, \citenamefont {Rodr\'{i}guez},\ and\
  \citenamefont {Johnson}}]{PRLart}%
  \BibitemOpen
  \bibfield  {author} {\bibinfo {author} {\bibfnamefont {O.~L.}\ \bibnamefont
  {Acevedo}}, \bibinfo {author} {\bibfnamefont {L.}~\bibnamefont {Quiroga}},
  \bibinfo {author} {\bibfnamefont {F.~J.}\ \bibnamefont {Rodr\'{i}guez}}, \
  and\ \bibinfo {author} {\bibfnamefont {N.~F.}\ \bibnamefont {Johnson}},\
  }\href {\doibase 10.1103/physrevlett.112.030403} {\bibfield  {journal}
  {\bibinfo  {journal} {Phys. Rev. Lett.}\ }\textbf {\bibinfo {volume} {112}},\
  \bibinfo {pages} {030403} (\bibinfo {year} {2014})}\BibitemShut {NoStop}%
\bibitem [{\citenamefont {Fuchs}\ \emph {et~al.}(2015)\citenamefont {Fuchs},
  \citenamefont {Ankerhold}, \citenamefont {Blencowe},\ and\ \citenamefont
  {Kubala}}]{Fuchs}%
  \BibitemOpen
  \bibfield  {author} {\bibinfo {author} {\bibfnamefont {S.}~\bibnamefont
  {Fuchs}}, \bibinfo {author} {\bibfnamefont {J.}~\bibnamefont {Ankerhold}},
  \bibinfo {author} {\bibfnamefont {M.}~\bibnamefont {Blencowe}}, \ and\
  \bibinfo {author} {\bibfnamefont {B.}~\bibnamefont {Kubala}},\ }\href@noop {}
  {\  (\bibinfo {year} {2015})},\ \Eprint
  {http://arxiv.org/abs/arXiv:1501.07841 [quant-ph]} {arXiv:1501.07841
  [quant-ph]} \BibitemShut {NoStop}%
\bibitem [{\citenamefont {Bastidas}\ \emph {et~al.}(2012)\citenamefont
  {Bastidas}, \citenamefont {Emary}, \citenamefont {Regler},\ and\
  \citenamefont {Brandes}}]{BastidasDicke}%
  \BibitemOpen
  \bibfield  {author} {\bibinfo {author} {\bibfnamefont {V.~M.}\ \bibnamefont
  {Bastidas}}, \bibinfo {author} {\bibfnamefont {C.}~\bibnamefont {Emary}},
  \bibinfo {author} {\bibfnamefont {B.}~\bibnamefont {Regler}}, \ and\ \bibinfo
  {author} {\bibfnamefont {T.}~\bibnamefont {Brandes}},\ }\href {\doibase
  10.1103/PhysRevLett.108.043003} {\bibfield  {journal} {\bibinfo  {journal}
  {Phys. Rev. Lett.}\ }\textbf {\bibinfo {volume} {108}},\ \bibinfo {pages}
  {043003} (\bibinfo {year} {2012})}\BibitemShut {NoStop}%
\bibitem [{\citenamefont {Altland}\ and\ \citenamefont
  {Haake}(2012)}]{Altland2}%
  \BibitemOpen
  \bibfield  {author} {\bibinfo {author} {\bibfnamefont {A.}~\bibnamefont
  {Altland}}\ and\ \bibinfo {author} {\bibfnamefont {F.}~\bibnamefont
  {Haake}},\ }\href {\doibase 10.1088/1367-2630/14/7/073011} {\bibfield
  {journal} {\bibinfo  {journal} {New J. Phys.}\ }\textbf {\bibinfo {volume}
  {14}},\ \bibinfo {pages} {073011} (\bibinfo {year} {2012})}\BibitemShut
  {NoStop}%
\bibitem [{\citenamefont {Alvermann}\ \emph {et~al.}(2012)\citenamefont
  {Alvermann}, \citenamefont {Bakemeier},\ and\ \citenamefont
  {Fehske}}]{Alvermann}%
  \BibitemOpen
  \bibfield  {author} {\bibinfo {author} {\bibfnamefont {A.}~\bibnamefont
  {Alvermann}}, \bibinfo {author} {\bibfnamefont {L.}~\bibnamefont
  {Bakemeier}}, \ and\ \bibinfo {author} {\bibfnamefont {H.}~\bibnamefont
  {Fehske}},\ }\href {\doibase 10.1103/physreva.85.043803} {\bibfield
  {journal} {\bibinfo  {journal} {Phys. Rev. A}\ }\textbf {\bibinfo {volume}
  {85}},\ \bibinfo {pages} {043803} (\bibinfo {year} {2012})}\BibitemShut
  {NoStop}%
\bibitem [{\citenamefont {Bakemeier}\ \emph {et~al.}(2013)\citenamefont
  {Bakemeier}, \citenamefont {Alvermann},\ and\ \citenamefont
  {Fehske}}]{Alvermann1}%
  \BibitemOpen
  \bibfield  {author} {\bibinfo {author} {\bibfnamefont {L.}~\bibnamefont
  {Bakemeier}}, \bibinfo {author} {\bibfnamefont {A.}~\bibnamefont
  {Alvermann}}, \ and\ \bibinfo {author} {\bibfnamefont {H.}~\bibnamefont
  {Fehske}},\ }\href {\doibase 10.1103/physreva.88.043835} {\bibfield
  {journal} {\bibinfo  {journal} {Phys. Rev. A}\ }\textbf {\bibinfo {volume}
  {88}},\ \bibinfo {pages} {043835} (\bibinfo {year} {2013})}\BibitemShut
  {NoStop}%
\bibitem [{\citenamefont {Furuya}\ \emph {et~al.}(1998)\citenamefont {Furuya},
  \citenamefont {Nemes},\ and\ \citenamefont {Pellegrino}}]{Furuya}%
  \BibitemOpen
  \bibfield  {author} {\bibinfo {author} {\bibfnamefont {K.}~\bibnamefont
  {Furuya}}, \bibinfo {author} {\bibfnamefont {M.~C.}\ \bibnamefont {Nemes}}, \
  and\ \bibinfo {author} {\bibfnamefont {G.~Q.}\ \bibnamefont {Pellegrino}},\
  }\href {\doibase 10.1103/physrevlett.80.5524} {\bibfield  {journal} {\bibinfo
   {journal} {Phys. Rev. Lett.}\ }\textbf {\bibinfo {volume} {80}},\ \bibinfo
  {pages} {5524} (\bibinfo {year} {1998})}\BibitemShut {NoStop}%
\bibitem [{\citenamefont {Song}\ \emph {et~al.}(2009)\citenamefont {Song},
  \citenamefont {Yan}, \citenamefont {Ma},\ and\ \citenamefont {Wang}}]{Song}%
  \BibitemOpen
  \bibfield  {author} {\bibinfo {author} {\bibfnamefont {L.}~\bibnamefont
  {Song}}, \bibinfo {author} {\bibfnamefont {D.}~\bibnamefont {Yan}}, \bibinfo
  {author} {\bibfnamefont {J.}~\bibnamefont {Ma}}, \ and\ \bibinfo {author}
  {\bibfnamefont {X.}~\bibnamefont {Wang}},\ }\href {\doibase
  10.1103/physreve.79.046220} {\bibfield  {journal} {\bibinfo  {journal} {Phys.
  Rev. E}\ }\textbf {\bibinfo {volume} {79}},\ \bibinfo {pages} {046220}
  (\bibinfo {year} {2009})}\BibitemShut {NoStop}%
\bibitem [{\citenamefont {Dicke}(1954)}]{Dicke}%
  \BibitemOpen
  \bibfield  {author} {\bibinfo {author} {\bibfnamefont {R.~H.}\ \bibnamefont
  {Dicke}},\ }\href {\doibase 10.1103/PhysRev.93.99} {\bibfield  {journal}
  {\bibinfo  {journal} {Phys. Rev.}\ }\textbf {\bibinfo {volume} {93}},\
  \bibinfo {pages} {99} (\bibinfo {year} {1954})}\BibitemShut {NoStop}%
\bibitem [{\citenamefont {Nataf}\ and\ \citenamefont {Ciuti}(2010)}]{Ciuti}%
  \BibitemOpen
  \bibfield  {author} {\bibinfo {author} {\bibfnamefont {P.}~\bibnamefont
  {Nataf}}\ and\ \bibinfo {author} {\bibfnamefont {C.}~\bibnamefont {Ciuti}},\
  }\href {\doibase 10.1038/ncomms1069} {\bibfield  {journal} {\bibinfo
  {journal} {Nature Comm.}\ }\textbf {\bibinfo {volume} {1}},\ \bibinfo {pages}
  {1} (\bibinfo {year} {2010})}\BibitemShut {NoStop}%
\bibitem [{\citenamefont {Viehmann}\ \emph {et~al.}(2011)\citenamefont
  {Viehmann}, \citenamefont {von Delft},\ and\ \citenamefont
  {Marquardt}}]{Marquadt}%
  \BibitemOpen
  \bibfield  {author} {\bibinfo {author} {\bibfnamefont {O.}~\bibnamefont
  {Viehmann}}, \bibinfo {author} {\bibfnamefont {J.}~\bibnamefont {von Delft}},
  \ and\ \bibinfo {author} {\bibfnamefont {F.}~\bibnamefont {Marquardt}},\
  }\href {\doibase 10.1103/physrevlett.107.113602} {\bibfield  {journal}
  {\bibinfo  {journal} {Phys. Rev. Lett.}\ }\textbf {\bibinfo {volume} {107}},\
  \bibinfo {pages} {113602} (\bibinfo {year} {2011})}\BibitemShut {NoStop}%
\bibitem [{\citenamefont {Baumann}\ \emph {et~al.}(2010)\citenamefont
  {Baumann}, \citenamefont {Guerlin}, \citenamefont {Brennecke},\ and\
  \citenamefont {Esslinger}}]{Baumann}%
  \BibitemOpen
  \bibfield  {author} {\bibinfo {author} {\bibfnamefont {K.}~\bibnamefont
  {Baumann}}, \bibinfo {author} {\bibfnamefont {C.}~\bibnamefont {Guerlin}},
  \bibinfo {author} {\bibfnamefont {F.}~\bibnamefont {Brennecke}}, \ and\
  \bibinfo {author} {\bibfnamefont {T.}~\bibnamefont {Esslinger}},\ }\href
  {\doibase 10.1038/nature09009} {\bibfield  {journal} {\bibinfo  {journal}
  {Nature}\ }\textbf {\bibinfo {volume} {464}},\ \bibinfo {pages} {1301}
  (\bibinfo {year} {2010})}\BibitemShut {NoStop}%
\bibitem [{\citenamefont {Hamner}\ \emph {et~al.}(2014)\citenamefont {Hamner},
  \citenamefont {Qu}, \citenamefont {Zhang}, \citenamefont {Chang},
  \citenamefont {Gong}, \citenamefont {Zhang},\ and\ \citenamefont
  {Engels}}]{DMEXPHamner}%
  \BibitemOpen
  \bibfield  {author} {\bibinfo {author} {\bibfnamefont {C.}~\bibnamefont
  {Hamner}}, \bibinfo {author} {\bibfnamefont {C.}~\bibnamefont {Qu}}, \bibinfo
  {author} {\bibfnamefont {Y.}~\bibnamefont {Zhang}}, \bibinfo {author}
  {\bibfnamefont {J.}~\bibnamefont {Chang}}, \bibinfo {author} {\bibfnamefont
  {M.}~\bibnamefont {Gong}}, \bibinfo {author} {\bibfnamefont {C.}~\bibnamefont
  {Zhang}}, \ and\ \bibinfo {author} {\bibfnamefont {P.}~\bibnamefont
  {Engels}},\ }\href {\doibase 10.1038/ncomms5023} {\bibfield  {journal}
  {\bibinfo  {journal} {Nature Comm.}\ }\textbf {\bibinfo {volume} {5}},\
  \bibinfo {pages} {4023} (\bibinfo {year} {2014})}\BibitemShut {NoStop}%
\bibitem [{\citenamefont {Baden}\ \emph {et~al.}(2014)\citenamefont {Baden},
  \citenamefont {Arnold}, \citenamefont {Grimsmo}, \citenamefont {Parkins},\
  and\ \citenamefont {Barrett}}]{DMEXPBaden}%
  \BibitemOpen
  \bibfield  {author} {\bibinfo {author} {\bibfnamefont {M.~P.}\ \bibnamefont
  {Baden}}, \bibinfo {author} {\bibfnamefont {K.~J.}\ \bibnamefont {Arnold}},
  \bibinfo {author} {\bibfnamefont {A.~L.}\ \bibnamefont {Grimsmo}}, \bibinfo
  {author} {\bibfnamefont {S.}~\bibnamefont {Parkins}}, \ and\ \bibinfo
  {author} {\bibfnamefont {M.~D.}\ \bibnamefont {Barrett}},\ }\href {\doibase
  10.1103/physrevlett.113.020408} {\bibfield  {journal} {\bibinfo  {journal}
  {Phys. Rev. Lett.}\ }\textbf {\bibinfo {volume} {113}},\ \bibinfo {pages}
  {020408} (\bibinfo {year} {2014})}\BibitemShut {NoStop}%
\bibitem [{\citenamefont {Klinder}\ \emph {et~al.}(2015)\citenamefont
  {Klinder}, \citenamefont {Ke{\ss}ler}, \citenamefont {Wolke}, \citenamefont
  {Mathey},\ and\ \citenamefont {Hemmerich}}]{PNAS}%
  \BibitemOpen
  \bibfield  {author} {\bibinfo {author} {\bibfnamefont {J.}~\bibnamefont
  {Klinder}}, \bibinfo {author} {\bibfnamefont {H.}~\bibnamefont {Ke{\ss}ler}},
  \bibinfo {author} {\bibfnamefont {M.}~\bibnamefont {Wolke}}, \bibinfo
  {author} {\bibfnamefont {L.}~\bibnamefont {Mathey}}, \ and\ \bibinfo {author}
  {\bibfnamefont {A.}~\bibnamefont {Hemmerich}},\ }\href {\doibase
  10.1073/pnas.1417132112} {\bibfield  {journal} {\bibinfo  {journal} {Proc.
  Natl. Acad. Sci. U. S. A.}\ }\textbf {\bibinfo {volume} {112}},\ \bibinfo
  {pages} {3290} (\bibinfo {year} {2015})}\BibitemShut {NoStop}%
\bibitem [{\citenamefont {Reslen}\ \emph {et~al.}(2005)\citenamefont {Reslen},
  \citenamefont {Quiroga},\ and\ \citenamefont {Johnson}}]{ReslenEPL}%
  \BibitemOpen
  \bibfield  {author} {\bibinfo {author} {\bibfnamefont {J.}~\bibnamefont
  {Reslen}}, \bibinfo {author} {\bibfnamefont {L.}~\bibnamefont {Quiroga}}, \
  and\ \bibinfo {author} {\bibfnamefont {N.~F.}\ \bibnamefont {Johnson}},\
  }\href {\doibase 10.1209/epl/i2004-10313-4} {\bibfield  {journal} {\bibinfo
  {journal} {Europhys. Lett.}\ }\textbf {\bibinfo {volume} {69}},\ \bibinfo
  {pages} {8} (\bibinfo {year} {2005})}\BibitemShut {NoStop}%
\bibitem [{\citenamefont {Vidal}\ and\ \citenamefont
  {Dusuel}(2006)}]{VidalDicke}%
  \BibitemOpen
  \bibfield  {author} {\bibinfo {author} {\bibfnamefont {J.}~\bibnamefont
  {Vidal}}\ and\ \bibinfo {author} {\bibfnamefont {S.}~\bibnamefont {Dusuel}},\
  }\href {\doibase 10.1209/epl/i2006-10041-9} {\bibfield  {journal} {\bibinfo
  {journal} {Europhys. Lett.}\ }\textbf {\bibinfo {volume} {74}},\ \bibinfo
  {pages} {817} (\bibinfo {year} {2006})}\BibitemShut {NoStop}%
\bibitem [{\citenamefont {Wang}\ \emph {et~al.}(2014)\citenamefont {Wang},
  \citenamefont {Wu}, \citenamefont {Yang}, \citenamefont {Jin}, \citenamefont
  {Lambert},\ and\ \citenamefont {Nori}}]{Wang2014}%
  \BibitemOpen
  \bibfield  {author} {\bibinfo {author} {\bibfnamefont {T.-L.}\ \bibnamefont
  {Wang}}, \bibinfo {author} {\bibfnamefont {L.-N.}\ \bibnamefont {Wu}},
  \bibinfo {author} {\bibfnamefont {W.}~\bibnamefont {Yang}}, \bibinfo {author}
  {\bibfnamefont {G.-R.}\ \bibnamefont {Jin}}, \bibinfo {author} {\bibfnamefont
  {N.}~\bibnamefont {Lambert}}, \ and\ \bibinfo {author} {\bibfnamefont
  {F.}~\bibnamefont {Nori}},\ }\href {\doibase 10.1088/1367-2630/16/6/063039}
  {\bibfield  {journal} {\bibinfo  {journal} {New J. Phys.}\ }\textbf {\bibinfo
  {volume} {16}},\ \bibinfo {pages} {063039} (\bibinfo {year}
  {2014})}\BibitemShut {NoStop}%
\bibitem [{\citenamefont {Jarrett}\ \emph {et~al.}(2007)\citenamefont
  {Jarrett}, \citenamefont {Olaya-Castro},\ and\ \citenamefont
  {Johnson}}]{OlayaCastroEPL}%
  \BibitemOpen
  \bibfield  {author} {\bibinfo {author} {\bibfnamefont {T.~C.}\ \bibnamefont
  {Jarrett}}, \bibinfo {author} {\bibfnamefont {A.}~\bibnamefont
  {Olaya-Castro}}, \ and\ \bibinfo {author} {\bibfnamefont {N.~F.}\
  \bibnamefont {Johnson}},\ }\href {\doibase 10.1209/0295-5075/77/34001}
  {\bibfield  {journal} {\bibinfo  {journal} {Europhys. Lett.}\ }\textbf
  {\bibinfo {volume} {77}},\ \bibinfo {pages} {34001} (\bibinfo {year}
  {2007})}\BibitemShut {NoStop}%
\bibitem [{\citenamefont {Eisert}\ \emph {et~al.}(2015)\citenamefont {Eisert},
  \citenamefont {Friesdorf},\ and\ \citenamefont {Gogolin}}]{eisertnature}%
  \BibitemOpen
  \bibfield  {author} {\bibinfo {author} {\bibfnamefont {J.}~\bibnamefont
  {Eisert}}, \bibinfo {author} {\bibfnamefont {M.}~\bibnamefont {Friesdorf}}, \
  and\ \bibinfo {author} {\bibfnamefont {C.}~\bibnamefont {Gogolin}},\ }\href
  {\doibase 10.1038/nphys3215} {\bibfield  {journal} {\bibinfo  {journal}
  {Nature Phys.}\ }\textbf {\bibinfo {volume} {11}},\ \bibinfo {pages} {124}
  (\bibinfo {year} {2015})}\BibitemShut {NoStop}%
\bibitem [{\citenamefont {Emary}\ and\ \citenamefont
  {Brandes}(2003{\natexlab{a}})}]{BrandesPRL}%
  \BibitemOpen
  \bibfield  {author} {\bibinfo {author} {\bibfnamefont {C.}~\bibnamefont
  {Emary}}\ and\ \bibinfo {author} {\bibfnamefont {T.}~\bibnamefont
  {Brandes}},\ }\href {\doibase 10.1103/PhysRevLett.90.044101} {\bibfield
  {journal} {\bibinfo  {journal} {Phys. Rev. Lett.}\ }\textbf {\bibinfo
  {volume} {90}},\ \bibinfo {pages} {044101} (\bibinfo {year}
  {2003}{\natexlab{a}})}\BibitemShut {NoStop}%
\bibitem [{\citenamefont {Emary}\ and\ \citenamefont
  {Brandes}(2003{\natexlab{b}})}]{BrandesPRE}%
  \BibitemOpen
  \bibfield  {author} {\bibinfo {author} {\bibfnamefont {C.}~\bibnamefont
  {Emary}}\ and\ \bibinfo {author} {\bibfnamefont {T.}~\bibnamefont
  {Brandes}},\ }\href {\doibase 10.1103/PhysRevE.67.066203} {\bibfield
  {journal} {\bibinfo  {journal} {Phys. Rev. E}\ }\textbf {\bibinfo {volume}
  {67}},\ \bibinfo {pages} {066203} (\bibinfo {year}
  {2003}{\natexlab{b}})}\BibitemShut {NoStop}%
\bibitem [{\citenamefont {Acevedo}(2015)}]{OscarTesis}%
  \BibitemOpen
  \bibfield  {author} {\bibinfo {author} {\bibfnamefont {O.~L.}\ \bibnamefont
  {Acevedo}},\ }\href@noop {} {\bibfield  {journal} {\bibinfo  {journal} {Ph.D.
  Thesis, Universidad de Los Andes}\ } (\bibinfo {year} {2015})}\BibitemShut
  {NoStop}%
\bibitem [{\citenamefont {Nielsen}\ and\ \citenamefont
  {Chuang}(2010)}]{Nielsen}%
  \BibitemOpen
  \bibfield  {author} {\bibinfo {author} {\bibfnamefont {M.~A.}\ \bibnamefont
  {Nielsen}}\ and\ \bibinfo {author} {\bibfnamefont {I.~L.}\ \bibnamefont
  {Chuang}},\ }\href@noop {} {\emph {\bibinfo {title} {Quantum Computation and
  Quantum Information}}}\ (\bibinfo  {publisher} {Cambridge University Press},\
  \bibinfo {year} {2010})\BibitemShut {NoStop}%
\bibitem [{\citenamefont {Alba}\ and\ \citenamefont
  {Heidrich-Meisner}(2014)}]{alba}%
  \BibitemOpen
  \bibfield  {author} {\bibinfo {author} {\bibfnamefont {V.}~\bibnamefont
  {Alba}}\ and\ \bibinfo {author} {\bibfnamefont {F.}~\bibnamefont
  {Heidrich-Meisner}},\ }\href {\doibase 10.1103/physrevb.90.075144} {\bibfield
   {journal} {\bibinfo  {journal} {Phys. Rev. B}\ }\textbf {\bibinfo {volume}
  {90}},\ \bibinfo {pages} {075144} (\bibinfo {year} {2014})}\BibitemShut
  {NoStop}%
\bibitem [{\citenamefont {Ma}\ \emph {et~al.}(2011)\citenamefont {Ma},
  \citenamefont {Wang}, \citenamefont {Sun},\ and\ \citenamefont
  {Nori}}]{QuaSpiSqu}%
  \BibitemOpen
  \bibfield  {author} {\bibinfo {author} {\bibfnamefont {J.}~\bibnamefont
  {Ma}}, \bibinfo {author} {\bibfnamefont {X.}~\bibnamefont {Wang}}, \bibinfo
  {author} {\bibfnamefont {C.}~\bibnamefont {Sun}}, \ and\ \bibinfo {author}
  {\bibfnamefont {F.}~\bibnamefont {Nori}},\ }\href {\doibase
  10.1016/j.physrep.2011.08.003} {\bibfield  {journal} {\bibinfo  {journal}
  {Phys. Rep.}\ }\textbf {\bibinfo {volume} {509}},\ \bibinfo {pages} {89}
  (\bibinfo {year} {2011})}\BibitemShut {NoStop}%
\bibitem [{\citenamefont {Law}\ \emph {et~al.}(2001)\citenamefont {Law},
  \citenamefont {Ng},\ and\ \citenamefont {Leung}}]{Transverse}%
  \BibitemOpen
  \bibfield  {author} {\bibinfo {author} {\bibfnamefont {C.~K.}\ \bibnamefont
  {Law}}, \bibinfo {author} {\bibfnamefont {H.~T.}\ \bibnamefont {Ng}}, \ and\
  \bibinfo {author} {\bibfnamefont {P.~T.}\ \bibnamefont {Leung}},\ }\href
  {\doibase 10.1103/physreva.63.055601} {\bibfield  {journal} {\bibinfo
  {journal} {Phys. Rev. A}\ }\textbf {\bibinfo {volume} {63}},\ \bibinfo
  {pages} {055601} (\bibinfo {year} {2001})}\BibitemShut {NoStop}%
\bibitem [{\citenamefont {Stobi\'{n}ska}\ \emph {et~al.}(2008)\citenamefont
  {Stobi\'{n}ska}, \citenamefont {Milburn},\ and\ \citenamefont
  {W\'{o}dkiewicz}}]{Milburn1}%
  \BibitemOpen
  \bibfield  {author} {\bibinfo {author} {\bibfnamefont {M.}~\bibnamefont
  {Stobi\'{n}ska}}, \bibinfo {author} {\bibfnamefont {G.~J.}\ \bibnamefont
  {Milburn}}, \ and\ \bibinfo {author} {\bibfnamefont {K.}~\bibnamefont
  {W\'{o}dkiewicz}},\ }\href {\doibase 10.1103/PhysRevA.78.013810} {\bibfield
  {journal} {\bibinfo  {journal} {Phys. Rev. A}\ }\textbf {\bibinfo {volume}
  {78}},\ \bibinfo {pages} {013810} (\bibinfo {year} {2008})}\BibitemShut
  {NoStop}%
\bibitem [{\citenamefont {Dowling}\ \emph {et~al.}(1994)\citenamefont
  {Dowling}, \citenamefont {Agarwal},\ and\ \citenamefont
  {Schleich}}]{WignerQubit1}%
  \BibitemOpen
  \bibfield  {author} {\bibinfo {author} {\bibfnamefont {J.~P.}\ \bibnamefont
  {Dowling}}, \bibinfo {author} {\bibfnamefont {G.~S.}\ \bibnamefont
  {Agarwal}}, \ and\ \bibinfo {author} {\bibfnamefont {W.~P.}\ \bibnamefont
  {Schleich}},\ }\href {\doibase 10.1103/PhysRevA.49.4101} {\bibfield
  {journal} {\bibinfo  {journal} {Phys. Rev. A}\ }\textbf {\bibinfo {volume}
  {49}},\ \bibinfo {pages} {4101} (\bibinfo {year} {1994})}\BibitemShut
  {NoStop}%
\bibitem [{\citenamefont {Vidal}\ and\ \citenamefont
  {Werner}(2002)}]{Vidalneg}%
  \BibitemOpen
  \bibfield  {author} {\bibinfo {author} {\bibfnamefont {G.}~\bibnamefont
  {Vidal}}\ and\ \bibinfo {author} {\bibfnamefont {R.~F.}\ \bibnamefont
  {Werner}},\ }\href {\doibase 10.1103/physreva.65.032314} {\bibfield
  {journal} {\bibinfo  {journal} {Phys. Rev. A}\ }\textbf {\bibinfo {volume}
  {65}},\ \bibinfo {pages} {032314} (\bibinfo {year} {2002})}\BibitemShut
  {NoStop}%
\bibitem [{\citenamefont {Giovannetti}\ \emph {et~al.}(2011)\citenamefont
  {Giovannetti}, \citenamefont {Lloyd},\ and\ \citenamefont
  {Maccone}}]{Metrology}%
  \BibitemOpen
  \bibfield  {author} {\bibinfo {author} {\bibfnamefont {V.}~\bibnamefont
  {Giovannetti}}, \bibinfo {author} {\bibfnamefont {S.}~\bibnamefont {Lloyd}},
  \ and\ \bibinfo {author} {\bibfnamefont {L.}~\bibnamefont {Maccone}},\ }\href
  {\doibase 10.1038/nphoton.2011.35} {\bibfield  {journal} {\bibinfo  {journal}
  {Nature Photon.}\ }\textbf {\bibinfo {volume} {5}},\ \bibinfo {pages} {222}
  (\bibinfo {year} {2011})}\BibitemShut {NoStop}%
\bibitem [{\citenamefont {Schnabel}\ \emph {et~al.}(2010)\citenamefont
  {Schnabel}, \citenamefont {Mavalvala}, \citenamefont {McClelland},\ and\
  \citenamefont {Lam}}]{MetrologyOPT}%
  \BibitemOpen
  \bibfield  {author} {\bibinfo {author} {\bibfnamefont {R.}~\bibnamefont
  {Schnabel}}, \bibinfo {author} {\bibfnamefont {N.}~\bibnamefont {Mavalvala}},
  \bibinfo {author} {\bibfnamefont {D.~E.}\ \bibnamefont {McClelland}}, \ and\
  \bibinfo {author} {\bibfnamefont {P.~K.}\ \bibnamefont {Lam}},\ }\href
  {\doibase 10.1038/ncomms1122} {\bibfield  {journal} {\bibinfo  {journal}
  {Nature Comm.}\ }\textbf {\bibinfo {volume} {1}},\ \bibinfo {pages} {121}
  (\bibinfo {year} {2010})}\BibitemShut {NoStop}%
\bibitem [{\citenamefont {Rey}\ \emph {et~al.}(2007)\citenamefont {Rey},
  \citenamefont {Jiang},\ and\ \citenamefont {Lukin}}]{Rey2007}%
  \BibitemOpen
  \bibfield  {author} {\bibinfo {author} {\bibfnamefont {A.~M.}\ \bibnamefont
  {Rey}}, \bibinfo {author} {\bibfnamefont {L.}~\bibnamefont {Jiang}}, \ and\
  \bibinfo {author} {\bibfnamefont {M.~D.}\ \bibnamefont {Lukin}},\ }\href
  {\doibase 10.1103/physreva.76.053617} {\bibfield  {journal} {\bibinfo
  {journal} {Phys. Rev. A}\ }\textbf {\bibinfo {volume} {76}},\ \bibinfo
  {pages} {053617} (\bibinfo {year} {2007})}\BibitemShut {NoStop}%
\bibitem [{\citenamefont {Braunstein}(2005)}]{QuInCo}%
  \BibitemOpen
  \bibfield  {author} {\bibinfo {author} {\bibfnamefont {S.~L.}\ \bibnamefont
  {Braunstein}},\ }\href {\doibase 10.1103/revmodphys.77.513} {\bibfield
  {journal} {\bibinfo  {journal} {Rev. Mod. Phys.}\ }\textbf {\bibinfo {volume}
  {77}},\ \bibinfo {pages} {513} (\bibinfo {year} {2005})}\BibitemShut
  {NoStop}%
\bibitem [{\citenamefont {Amico}\ \emph {et~al.}(2008)\citenamefont {Amico},
  \citenamefont {Osterloh},\ and\ \citenamefont {Vedral}}]{VedralE}%
  \BibitemOpen
  \bibfield  {author} {\bibinfo {author} {\bibfnamefont {L.}~\bibnamefont
  {Amico}}, \bibinfo {author} {\bibfnamefont {A.}~\bibnamefont {Osterloh}}, \
  and\ \bibinfo {author} {\bibfnamefont {V.}~\bibnamefont {Vedral}},\ }\href
  {\doibase 10.1103/revmodphys.80.517} {\bibfield  {journal} {\bibinfo
  {journal} {Rev. Mod. Phys.}\ }\textbf {\bibinfo {volume} {80}},\ \bibinfo
  {pages} {517} (\bibinfo {year} {2008})}\BibitemShut {NoStop}%
\bibitem [{\citenamefont {Walls}\ and\ \citenamefont
  {Milburn}(2008)}]{quantumoptics}%
  \BibitemOpen
  \bibfield  {author} {\bibinfo {author} {\bibfnamefont {D.}~\bibnamefont
  {Walls}}\ and\ \bibinfo {author} {\bibfnamefont {G.~J.}\ \bibnamefont
  {Milburn}},\ }\href@noop {} {\emph {\bibinfo {title} {Quantum Optics}}}\
  (\bibinfo  {publisher} {Springer},\ \bibinfo {year} {2008})\BibitemShut
  {NoStop}%
\bibitem [{\citenamefont {Breuer}\ and\ \citenamefont
  {Petruccione}(2007)}]{breuer}%
  \BibitemOpen
  \bibfield  {author} {\bibinfo {author} {\bibfnamefont {H.~P.}\ \bibnamefont
  {Breuer}}\ and\ \bibinfo {author} {\bibfnamefont {F.}~\bibnamefont
  {Petruccione}},\ }\href@noop {} {\emph {\bibinfo {title} {The Theory of Open
  Quantum Systems}}}\ (\bibinfo  {publisher} {Oxford University Press},\
  \bibinfo {year} {2007})\BibitemShut {NoStop}%
\end{thebibliography}%

\end{document}